# Constructing a Nanopipette-based DNA Electro-Mechanical Device

*Cengiz J. Khan, Oliver J. Irving, Rand A. Al-Waqfi, Giorgio Ferrari, Tim Albrecht\**

University of Birmingham, School of Chemistry, Edgbaston Campus, Birmingham B15 2TT, United Kingdom

Solid-state nanopore and nanopipette sensors are powerful devices for the detection, quantification and structural analysis of biopolymers such as DNA and proteins, especially in carrier-enhanced resistive-pulse sensing. However, hundreds of different molecules typically need to be sampled from solution and analysed to obtain statistically robust information. This limits the applicability of such sensors and complicates associated workflows. Here, we present a new strategy to trap DNA structures in the sensing region of a nanopipette through end functionalisation and nanoparticle capping. We develop a robust set of descriptors to characterise the insertion and presence of nanoparticle-DNA constructs in the nanopipette tip and, furthermore, show that they remain mobile and responsive to external electric fields over extended periods of time. This allows for repeated readout of the same DNA structure and could enable new applications for such sensors, for example in flow and in confined environments.

Nanopore sensors are powerful tools for the study of small molecule transport through protein nanochannels, the detection of nanoparticles, DNA, RNA as well as proteins. They have transformed DNA and RNA sequencing, with interesting prospects towards protein sequencing as well.[1-5] Such platforms are also compatible with various detection methodologies, including electric, fluorescence, Raman, electrochemical and quantum tunnelling.[6-10] To this end, electric readout is of particular interest for point-of-care applications, as the small footprint, compatibility with microfluidic sample processing and device miniaturization are advantageous features in this regard.[11-13] The basic operating principle is relatively simple: a "resistive pulse" sensor is typically composed of two electrolyte-filled compartments, connected through a single, nanometer-sized pore or channel and equipped with a suitable, non-polarisable electrode in each one. Application of a bias voltage results in an electric current through the cell, and if the nanopore is the dominant source of resistance, the potential drop, and hence the electric field in the pore channel can be large. The latter can lead to the translocation of individual DNA, proteins or particles in solution, and if the pore resistance is sufficiently altered during this process, such an event may be detected directly via the measured current. More detailed analysis of current-time signature, for example in terms of duration, magnitude and sub-structure, can provide a wealth of information about the analyte in question.[14-17] This "richness" in information opens up interesting new avenues for bioanalytical applications as well. For example, in carrier-enhanced nanopore sensing, long, kilobase pair (kbp) DNA is functionalized with specific capture probes, such as antibodies, aptamers or oligonucleotides, in well-defined locations.[18-20] Upon incubation with analytes of interest, subsequent translocation of the carrier DNA can reveal the binding state of each capture probe, thereby confirm the presence of an analyte and provide an estimate of their concentrations.[19] Given the ability of typical nanopore sensors to resolve structural features on the carrier DNA that are less than 100 nm apart, this allows for multiplexed detection on a single carrier.[19, 21] Multiplexing capabilities can be further enhanced by mixing different carriers, as long as they are distinguishable, for example by length or by barcoding.[21] Finally, it has been noted that carrier-enhanced sensing may facilitate the detection in more complex mixtures, because the translocation signature of the DNA carrier may be used to isolate events of interest.[22, 23]

Typically, several hundred translocation events need to be recorded to build up a sufficiently robust statistical basis.[14-16, 24] These are based on different molecules or particles that are "lost" following the translocation through the pore, unless they are recaptured by fast bias reversal.[16, 25, 26]

This raises a more fundamental question, namely, whether a functional DNA carrier could be permanently trapped in the sensing region. In this way, sample incubation (target binding) and sensing process could be integrated, and repeated readout of the same carrier, for example by applying oscillatory electric fields, could be used to extract bioanalytical information. Such a platform could also facilitate measurements in flow, particularly for nanopipettes, in confined spaces, such as individual cells.[7, 27, 28]

Here, we demonstrate the step-by-step fabrication of such a nanoelectromechanical device (NEMD), by combining DNA engineering, nanoparticle chemistry and electrophoretically driven assembly in a nanopipette, fig. 1. We characterize the assembly systematically, with a combination of biochemical, structural, optical and electrical methodologies, and demonstrate the stable formation as well as the capability for bidirectional transport of the trapped DNA structure. As a key conceptual advance, our findings pave the way for a new bioanalytical device platform that could have a substantial impact on electric single-molecule sensing and its applications.

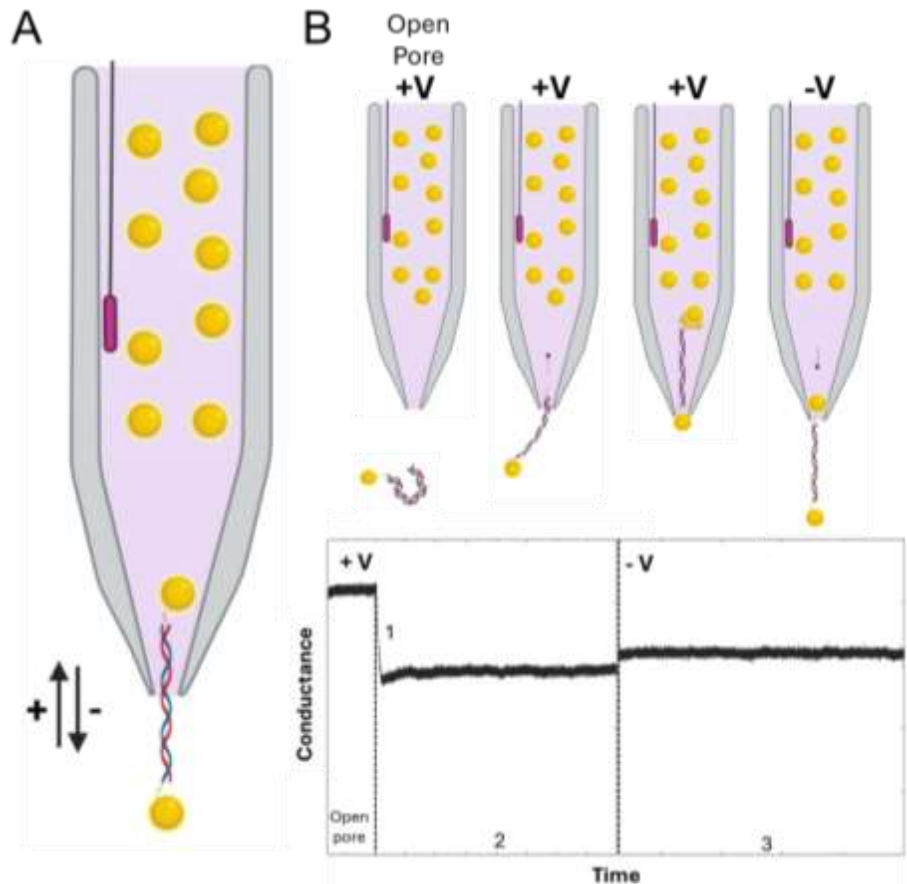

**Figure 1**. A) Illustration of the NP/DNA/NP construct trapped in a nanopipette. The application of a bias voltage can drive the structure in or out of the pipette, approximately over the length of the DNA. The NPs are larger than the inner pore diameter of the nanopipette, preventing the escape of the trapped structure B) Step-by-step assembly (top) and pore conductance as a function of time (bottom). Initially, the conductance is at the open-pore value. The DNA is modified with an azide group at one end and biotinyl at the other, allowing for orthogonal binding of monofunctionalised DBCO- and streptavidin-modified NPs, respectively. A NP/DNA construct is electrophoretically driven into the pipette ("1") and arrested as the NP is unable to translocate. Binding of the counter particle on the inside of nanopipette completes the NP/DNA/NP construct ("2"). The open-pore conductance is not recovered upon bias reversal ("3"), suggesting that the structure can no longer be ejected.

Our strategy is based on the idea that the binding of metallic nanoparticles to end-functionalised DNA can provide a "stopper" that prevents the escape of the DNA construct from the sensing region of the nanopipette. In our experiments, we used gold nanoparticles (AuNPs) that had a diameter substantially larger than the inner diameter of the nanopipette tip (~ 2:1). Preformed NP/DNA constructs were guided into the nanopipette using externally applied electric fields, an approach inspired by previous work on DNA Origami nanopores.[29] The second, "counter" particle was made available on the inside of the nanopipette, such that the formation of the complete NP/DNA/NP construct could only occur with the inserted NP/DNA complex, as illustrated in fig.1. The different steps in this process were monitored in real-time using electrical recordings.

We start by discussing the preparation and characterisation of the individual device components, namely the DNA, the NPs, as well as their complexes, fig. 2. Further experimental details can be found in section S1 of the supplementary information (SI). 5 kbp DNA was prepared using PCR amplification with primers carrying azide- and biotinylated groups at the respective 5' end. The final DNA product thus featured two orthogonal binding groups for DBCO- (dibenzocyclooctyne) and streptavidin-modified gold nanoparticles (core diameter: 40 ± 2 nm, Nanopartz, Loveland/USA). Importantly, both types of particles predominantly feature a single binding site, according to the manufacturer's specifications, thereby facilitating the specific formation of DNA/particle constructs and reducing the probability of oligomerization. Subsequently, different building blocks of the final NP/DNA/NP design were initially prepared in free solution and then characterized by gel electrophoresis (1% agarose, 80 V, 45 minutes), panel A. Lanes furthest to the left and right contain DNA ladder (GeneRuler™ DNA Ladder Mix, Thermo Scientifc™).

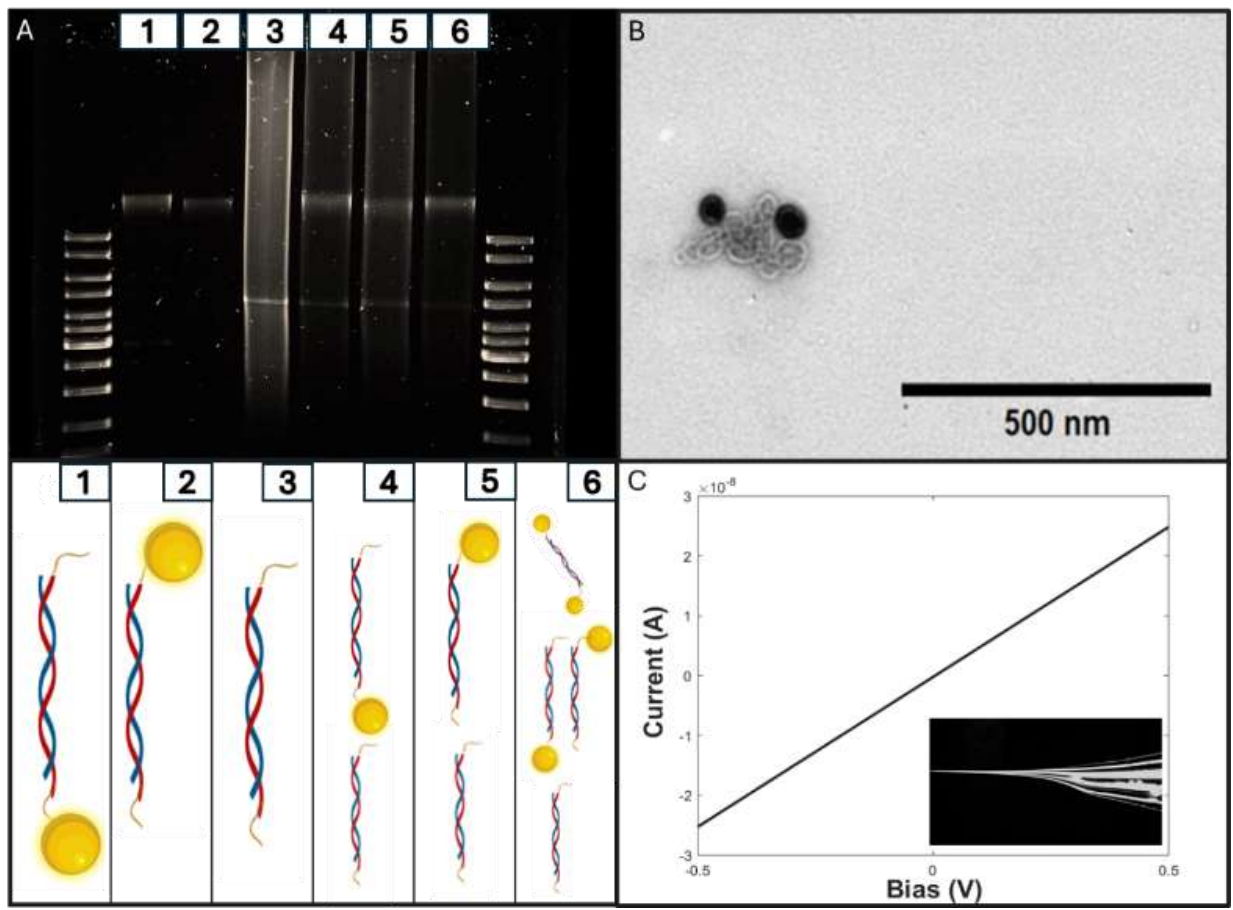

**Figure 2**. Characterisation of the components used to build the dumbbell device a) Agarose gel electrophoresis (1% agarose, 80V, 45 minutes). Sample 1: DBCO-AuNP bound to 5kbp DNA, Sample 2: Streptavidin-AuNP bound to 5kbp, Sample 3: unmodified 5kbp DNA, Sample 4: unpurified reaction mixture of DBCO-AuNP and 5kbp DNA, Sample 5: unpurified reaction mixture Streptavidin-AuNP and Sample 6: unpurified reaction mixture of all components. A schematic of agarose gel well contents can be seen below panel A. B) TEM image of coiled 5 kbp DNA bound to two gold nanoparticles. C) Nanopipette IV characteristic trace with optical image of the taper (inset).

Lanes 1 and 2 show the DNA conjugated with streptavidin- and azide-modified nanoparticles, respectively, after purification by centrifugation (5000 rpm, 10 minutes). A single, well-defined band is visible in each case, nominally at DNA lengths > 10 kbp. Lane 3 contains the raw PCR product, with the 5 kbp band clearly visible, and lanes 4 and 5 the same samples as in lanes 1 and 2, but before the removal of unbound DNA. Note that the bands for both 5 kbp DNA and conjugate are visible in those lanes. The significant reduction in mobility upon binding of the first particle, regardless of which one, was surprising given the small effect of the particle on the mass and overall charge of the construct. This suggests a more complex interaction between the NP/DNA construct and the gel matrix, as has indeed been observed previously.[30] The sample produced after reacting the DNA with both particles is shown in lane 6. While a faint band corresponding to unreacted 5 kbp DNA still appears, the dominant band is at a similar position as the NP/DNA constructs (lanes 1+2). This could either mean that the second particle was not bound, or not in insufficient quantities, or that the mobility of the full NP/DNA/NP construct is indeed not significantly different from the NP/DNA complex. We regard the former as unlikely, given the effective binding of each individual particle to the DNA, while the latter would be broadly in line with the notion that the presence of NPs limits the mobility of the construct through the gel matrix, *vide supra*. Hence, based on the gel electrophoresis data, we conclude that NP/DNA binding occurs and that the full NP/DNA/NP construct can indeed be formed, even if binding of the second particle does not result in a further reduction of the gel mobility. We note that similar observations have been made by Pelegrino et al. for gold nanoparticle-DNA conjugates of varying DNA lengths, lending further support to our interpretation.[31] Our gel electrophoresis results furthermore confirm that the NPs themselves possess a (small) positive charge (not shown), in accordance with the manufacturer's specification, suggesting that the mobility of the NP/DNA(/NP) constructs is dominated by the DNA.

Further support for the successful formation of the NP/DNA/NP construct is provided by TEM imaging, panel B. In this example, both the DNA and the NPs are well-resolved, and suggesting successful binding of both NPs to the DNA. More examples of TEM imaging of NP/DNA/NP mixtures are shown in fig. S1 in the SI.

Nanopipettes were prepared as reported previously, while the pulling parameters were optimized such that the pore diameter was approximately 20 nm, see section S4 in SI.[14, 15] Accordingly, panel C shows the current-voltage (IV) trace of a representative nanopipette chosen from a batch of 10 prepared under the same conditions. For each pipette, the pore conductance, $G_{pore}$, was determined from the slope of the IV trace between -0.5 V and +0.5 V, with an average of 43.5 nS and a standard deviation of 14.0 nS. $G_{pore}$ was then used to estimate the (inner) pore diameter, $d_i$, of the nanopipette (at the tip), based on eq. 1 in the SI. Hence, for the pipette shown in panel D, we obtained $G_{pore}$ = 47 nS and $d_i$ = 24 nm. An optical image is included in the inset, for reference.

Having prepared and characterised the individual building blocks for the device, the assembly of the NEMD was attempted next. For this purpose, the pre-prepared DBCO-NP/DNA complex was provided on the outside of the nanopipette, while the streptavidin-modified NP (strep-NP) was simultaneously present on the inside (electrolyte: 4 M LiCl; concentration of NP/DNA conjugate: ~28 pM) A negative $V_{bias}$ of -0.8 V was then applied to drive the NP/DNA complex into the nanopipette, in line with gel mobility results noted above.

Our expectation was that once the DNA part has entered the pore, the structure is arrested when the NP reaches the pore entrance, while a sufficiently large applied (negative) $V_{bias}$ effectively prevents escape in the opposite direction (noting that it is nevertheless subject to Brownian motion). This configuration would allow sufficient time for the strep-NP to bind, thus completing the formation of the NP/DNA/NP construct. Accordingly, we would expect the trapping of the structure to lead to a sustained reduction in the pore conductance that is furthermore maintained upon bias reversal (since it can no longer be ejected).

To monitor the above processes in real time, we take advantage of the fact that our setup features two output channels, [14-16] broadly speaking one containing slow (frequency components below $f_{DC}$ ~ 7 Hz, "DC channel") and the other one fast current modulations (frequency components higher than $f_{DC}$ up to about 2 MHz, "AC channel"). However, a detailed analysis reveals a more complex relationship between the nature of the input current modulation and the responses of the individual output channels, for results from circuit simulations and further details see section S9 in the SI. Briefly, short-lived and transient DNA translocation events with a characteristic time $\tau << 1/2\pi f_{DC}$ typically produce an approximately rectangular pulse in the input current. This is reflected in a similar response in the AC channel, while the DC channel remains unchanged. On the other hand, an insertion event of the kind discussed above is expected to produce a step-like change in the input current, provided the NP/DNA complex remains trapped in the nanopore. Initially, this results in a current modulation in the AC channel, which subsequently returns to its zero-mean value on a timescale of $1/2\pi f_{DC}$. In parallel, the DC channel evolves from the steady-state current value pre-insertion to a new one, representative of the nanopore with the trapped structure in place. In other words, insertion events become apparent from transient events in the AC channel and a concurrent, step-like change in the DC channel.

We have therefore expanded the analysis pipeline for the translocation data. In addition to the event magnitude $\Delta I_e$ and duration $\tau$ for events detected in the AC channel, we also determine the change in average $I_{DC}$, $\Delta I_{DC}$, and in the AC channel noise, $\Delta\sigma_{AC}$, before and after an event (based on 100 ms time windows before an event start and after the event ends in the AC channel), cf. section S5 in the SI for more details.

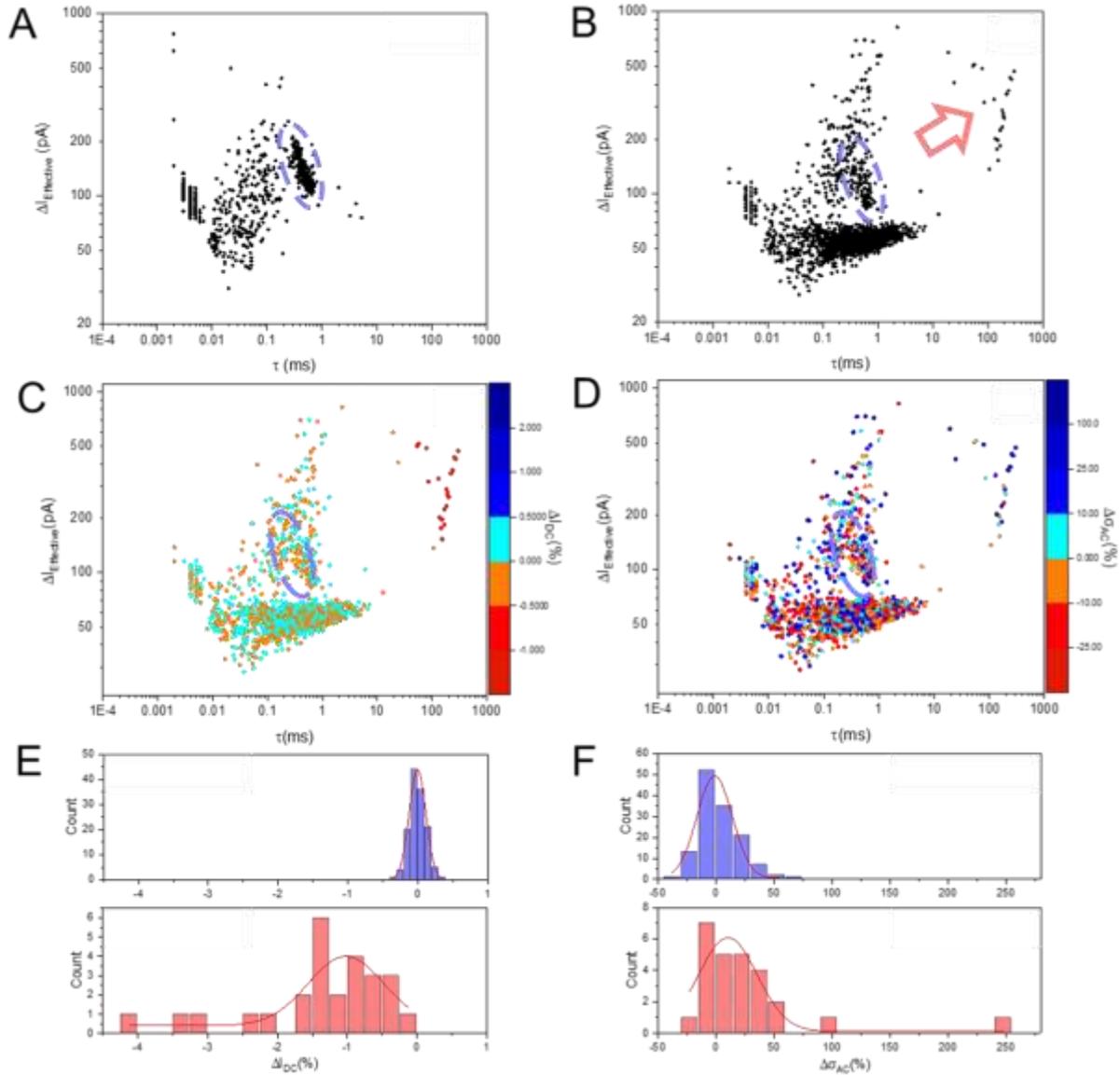

**Figure 3.** A) Translocation of the PCR solution with (end-functionalised) 5 kbp DNA as main product as well as shorter byproducts ($V_{bias}$ = -0.8 V; 4 M LiCl, $d_i$ = 19 nm). Very short and low-amplitude events are most likely due to electric noise, while translocation of PCR fragments results in a diffuse cluster with τ below approximately 0.1 ms. Translocation of the 5 kbp DNA results in a well-defined cluster (dashed ellipse) B) A similar experiment, but with NP/DNA complex on the outside and strep-NP on the inside of the nanopipette ($d_i$ = 24 nm). Results are broadly comparable with those in A), but with important differences, see main text for further discussion. The broad cluster < 70 pA is the result of baseline fluctuations. A new cluster emerges at $\tau \approx$ 100 ms and $\Delta I_e$ > 100 pA (red arrow), which we show corresponds to the insertion of NP/DNA complexes. C) Same dataset as in B, but with $\Delta I_{DC}$ colour-coded (in %). Conventional translocation of dsDNA results in an average $\Delta I_{DC} \approx 0$ (see histograms in panel E, top), while insertion events lead to a systematic decrease in $I_{DC}$ of about 1% on average (bottom). D) Same dataset as in B, but with $\Delta\sigma_{AC}$ colour-coded (in %). Conventional translocation of DNA results in $\Delta\sigma_{AC} \approx 0$ on average, while insertion events lead $\Delta\sigma_{AC} \approx$ 20% (panel F).

Use of the former is justified by the above considerations, namely that trapping of the NP/DNA construct should lead to a sustained reduction in $I_{DC}$ ($\Delta I_{DC}$ < 0). Choosing the latter was motivated by previous observations that dynamic processes and charge redistribution in the sensing region of the nanopore can lead to an increase in the current noise.[32, 33] Our hypothesis was therefore that trapping of the NP/DNA constructs could result in $\Delta\sigma_{AC}$ > 0, while transient occupation of the sensing region (such as for conventional translocation events) should leave the noise level largely unchanged ($\Delta\sigma_{AC} \approx 0$). This expectation was indeed borne out, as we show below. The noise characteristics of the DC channel were less relevant in this context, as it only captures slow modulations in the input current, as noted above. We show the results of such an analysis as scatter plots of $\Delta I_e$ vs. $\tau$ in fig. 3 ($V_{bias}$ = -0.8 V; 4 M LiCl, $d_{in}$ = 19 nm).

As noted above, the main product of the PCR reaction was 5 kbp DNA, functionalised at each end with an azide and a biotin group, respectively, in preparation for orthogonal NP coordination at each end. Both modifications are small and unlikely to obstruct translocation of the DNA through the nanopipette. Accordingly, the scatter plot in panel A shows a distinct cluster of translocation events at approximately $\tau \approx 0.5$ ms and $\Delta I_e \approx 100$-$200$ pA (dashed ellipse). Further analysis of the events in this cluster reveals that those due to linear translocation of DNA are primarily found towards the bottom right of this cluster (longer $\tau$, lower $\Delta I_e$), while folded events dominate towards the top left (shorter $\tau$, larger $\Delta I_e$), as is well-known based on previous work.[14-16] A more diffuse cluster of events is found between approximately $0.01 < \tau < 0.5$ ms and $50$ pA $< \Delta I_e < 200$ pA, which is absent in purified DNA samples (not shown).[14-16, 34] We therefore believe that these events arise from the translocation of DNA byproducts from the OCR reaction. This is in accordance with our gel electrophoresis data, cf. fig 2A, where smearing due to small non-specific DNA fragments is apparent. Finally, at $\tau < 0.01$ ms, a distinct cluster due to electric noise emerges. We note that there are essentially no events with $\tau > 1$ ms in this dataset.

We then performed the same experiment with a similar pipette ($d_i = 24$ nm), but now with the products from the NP/DNA assembly reaction on the outside and the respective counter (strep-modified) NP on the inside of the nanopipette. The scatter plot from the analysis of the translocation data, panel B, looks in some respects similar to the one in panel A, with a distinct cluster from DNA translocation (dashed ellipse), shorter events from PCR side products as well as electric noise at $\tau < 0.01$ ms. However, there are also important differences. Firstly, we note the dense cluster occurring at low magnitude ($< 70$ pA) over a wide range of $\tau$ values $0.01$ ms $< \tau < 10$ μσ), which we attribute to baseline fluctuations, possibly due to the initial equilibration of the sensor. Secondly, a small number of events occurs at $0.1$ ms $< \tau < 1$ ms with $\Delta I_e > 200$ pA, which is similar in duration to the 5 kbp DNA PCR product, but significantly larger in magnitude.

Their shapes vary, see fig. S5c in the SI for representative examples. We cannot rule out that some of those result from the translocation of NP/DNA complexes, where the NP is too small to be trapped. On the other hand, in TEM imaging studies the smallest particle diameter we have observed was ≈ 34 nm, cf. fig. S2, thus significantly larger than the pore diameter (here: $d_i$ ≈ 24 nm). Statistically, it would therefore be highly unlikely to capture a NP/DNA complex with a particle diameter < 24 nm. However, based on the large variance in event shapes and the relatively large event magnitude, we suggest that events in this cluster result from the insertion and subsequent, rapid dissociation of the NP/DNA complex. Thirdly, there is a new but distinct group of events with $\tau$ > 100 ms and 150 pA < $\Delta I_e$ < 500 pA (N ≈ 26, red arrow), i.e. with characteristic times two orders of magnitude larger than conventional DNA translocation events as well as increased $\Delta I_e$. These are unlikely conventional translocation events of DNA.

In order to better understand their origin, we subsequently show the same data as in B, but now with data points colour-coded according to $\Delta I_{DC}$ and $\Delta \sigma_{AC}$ in panels C and D, respectively. Even by visual inspection, it becomes clear that the two event classes – conventional DNA translocation and the long-lived ones at $\tau$ > 100 ms – indeed show very different behaviour, in terms of $\Delta I_{DC}$ and $\Delta \sigma_{AC}$. While conventional DNA translocation events do not feature systematic changes in $\Delta I_{DC}$ or $\Delta \sigma_{AC}$, see histograms in panels E and F, said long-lived events are qualitatively different. They are characterised by an average $\Delta I_{DC}$ ≈ -1% (decrease in $I_{DC}$ post event) and $\Delta \sigma_{AC}$ ≈ +20% (increase in $\sigma_{AC}$ post event), consistent with our expectations for NP/DNA insertion and trapping, thereby supporting our interpretation of the physical origin of these events. Those findings are also reminiscent of our recent work, where we found that (reversible) clogging of the nanochannel with 48.5 kbp DNA can lead to a concomitant significant decrease in $I_{DC}$ and an increase $\sigma_{AC}$, suggesting that the presence of DNA in the sensing region could be responsible for both.[16] In the present case, this would suggest not only that the DNA-NP structure has been inserted successfully, but also that $\sigma_{AC}$ may indeed be used as a proxy for its presence.

To conclude with some final observations, we have also found a marked, approximately fourfold decrease in the translocation frequency of 5 kbp DNA, comparing an open nanopipette with one with a trapped NP/DNA/NP structure, cf. fig. S10 in the SI. This is in line with the notion that while trapping does not completely block the pore entrance, it decreases its effective area and thus reduces the translocation frequency (of DNA). In some instances, insertion events featured multiple characteristic and transient "dips" in the AC channel current, possibly indicating consecutive interactions between the particle and the nanopipette opening. We show several examples in fig. S5d of the SI, but abstain from further, more detailed analysis, due to the limited size of the dataset.

More importantly, however, to demonstrate successful binding of the NP/DNA complex to the strep-NP counter particle on the inside of the nanopipette, we investigated the electric characteristics of the sensor post insertion at different bias polarities, as illustrated in fig. 4 A. Our hypothesis here is that the successfully formed NP/DNA/NP construct would no longer be ejected from the nanopipette under reverse bias ($V_{bias} > 0$), and that $G_{pore}$ would remain at a value corresponding to the occupied state of the pipette.

This is indeed borne out by the results summarised in fig. 4 B, which shows the probability distributions of $I_{DC}$ based on 10 s recording time for the first file of each $V_{bias}$ step (lasting 1000 s each), from "open pore" to $V_{bias}$ = -0.8 V, +0.8 V, -0.6 V and finally +0.6 V.

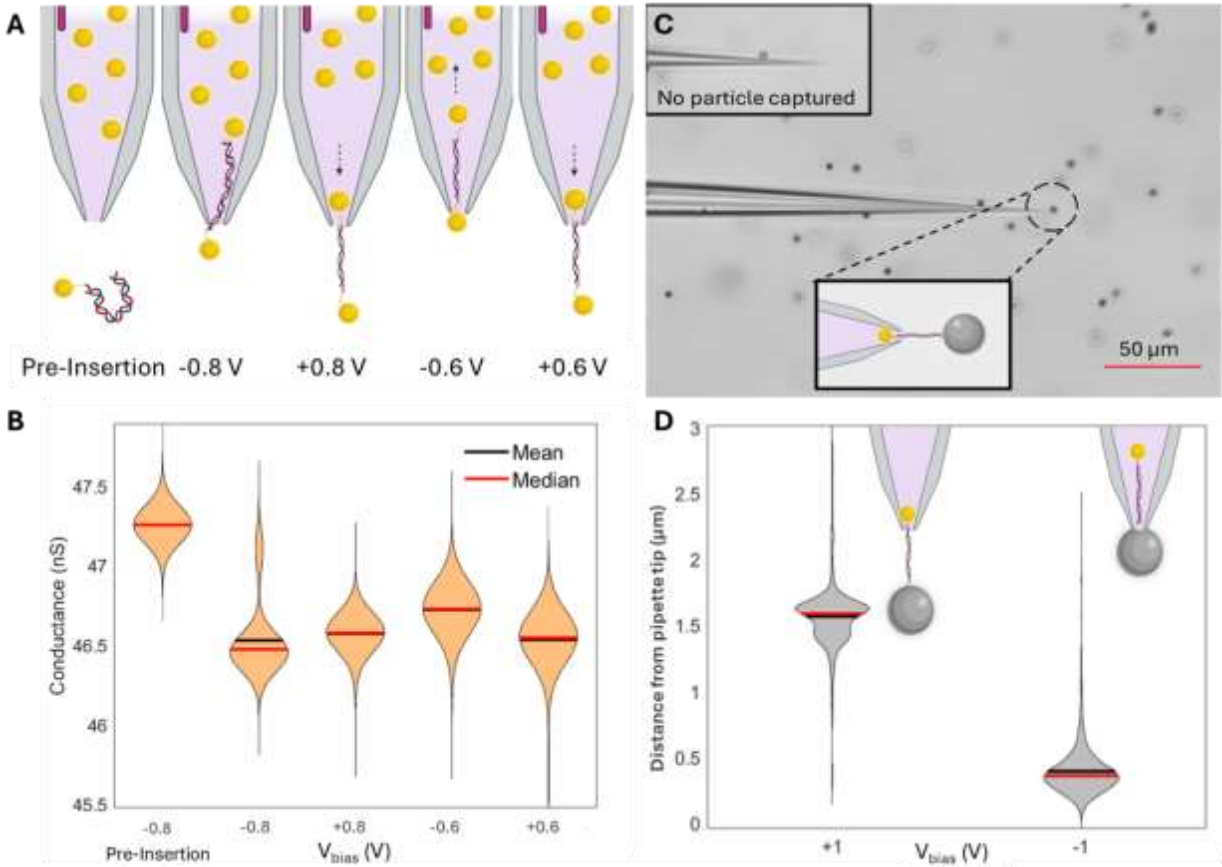

**Figure 4** A) Schematic illustrating the stepwise assembly of the NP/DNA/NP structures and their behavior under bias reversal. All experiments in 4 M LiCl B) Violin plots of the conductance for the different stages, from the open-pore (~47.2 nS, "pre-insertion") to full assembly under bias reversal. NP-DNA insertion causes a conductance drop (to 46.5 nS), which remains stable under successive bias reversals (+0.8 V, -0.6 V, +0.6 V) over approximately 3000 s. C) Optical image of a nanopipette with a trapped NP/DNA/MP construct (inset: pipette pre-insertion). The MP position is tracked during the different bias conditions (+1 V and -1 V), to determine the tether length and stability of the construct. D) Violin plots showing bead displacement measured from the pipette tip at +1 V (pushed outward) and -1 V (pushed inward). The distributions indicate a tether length of 1.2 μm. Insets: schematic depicting bead positions under bias reversal.

Note that the drop from the "pre-insertion" value (~47.2 nS) to the blocked state (~46.5 nS) occurs during the early stages of step 2 (-0.8 V), leading to a bimodal probability distribution during this step. $V_{bias}$ reversal to +0.8 V (step 3), however, does not lead to a recovery of the open pore value and $G_{pore}$ remains low at ~46.5 nS, indicating that the DNA has not been ejected. Correspondingly, $G_{pore}$ remains virtually unchanged in steps 3 ($V_{bias}$ = -0.6 V) and 4 ($V_{bias}$ = +0.6 V), suggesting that the NP/DNA/NP construct has indeed formed successfully and remains stable under applied bias for over ~3000 seconds in total. Finally, we provide further evidence that this is the case by replacing the outside NP by magnetic, streptavidin-modified polystyrene microparticles (MP, diameter: 1 μm; Dynabeads™, Invitrogen) that can be tracked directly using optical microscopy, cf. fig. 4 C and Methods for further details). The DBCO-NP is now present inside the nanochannel.

Accordingly, at $V_{bias}$ = -1 V, the DNA/MP complex is driven into the pipette tip and particle tracking analysis reveals that the MP resides very close to the tip end. Upon bias reversal to +1 V, the DNA is moved out of the pipette by approximately 1.2 ± 0.25 μm, based on the MP position, but does not altogether leave the tip region, due to successful tethering to the DBCO-NP on the inside of the pipette.

We note that the observed tether length is somewhat shorter than the full, linear length of 5 kbp DNA, which for B-DNA is approximately 1.7 μm.[28] Overall, this is not unexpected however, as there are several factors that could contribute to this effect. Firstly, the inside NP may not be able to reach the very end of the nanochannel, due to its small opening angle, thereby shortening the part of the tether that is outside of the pipette tip. Secondly, as the local electric field outside the nanopipette rapidly decreases with distance from the pore opening,[35, 36] the DNA tether may no longer be fully extended and (partially) refold into a more globular structure, as observed in free solution.[37, 38] Thirdly, considering the geometry of our setup and the size of the MP, gravitational effects could lead to downward bending of the tether, leading to an apparent shortening of the tether during imaging. Future studies may shed light into the relative contributions of the factors, but this does not take away from the key findings, namely that the assembly protocol successfully generates NP/DNA/NP and NP/DNA/MP structures trapped in the nanopipette and that these nevertheless remain mobile and stable at rather large biases in both polarities.

Conclusions

We have demonstrated a robust nanopore-based strategy for the stepwise assembly and permanent trapping of DNA-gold nanoparticle dumbbell structures using quartz nanopipettes with sub-30 nm apertures. By taking advantage of electric field-driven assembly of orthogonally functionalised NPs with end-labelled 5 kbp DNA, we achieved controlled, multi-step formation of NP/DNA/NP constructs directly within the nanopore. Real-time electrical recordings revealed distinct NP/DNA insertion events with sustained reduction in the pore current and increased noise. Voltage-switching experiments confirmed the stable formation of these structures, which was further supported by complementary optical measurements involving microparticles. The latter not only validated their mechanical integrity but also provided direct proof that the trapped structures nevertheless remain mobile and responsive to external electric fields. To this end, tether lengths were in line with expectation for 5 kbp DNA under the conditions used. Employing established DNA assembly and modification techniques, the DNA building blocks may be modified to include capture probes for specific biomolecular targets, while the ability to repeatedly transport such functionalised NP/DNA/NP structures through the sensing region of the nanopore may enable high-fidelity electric readout of target binding. Together, these findings establish a path to a novel, nanoelectromechanical sensing platform that may enable new modes of dynamic sensing and molecular interrogation, including in confined geometries and flow-based environments.

## ASSOCIATED CONTENT

**Supporting Information**.

The following files are available free of charge, containing further details on the synthesis and preparation of the DNA and the nanoparticle-DNA complexes; TEM imaging studies; nanopipette fabrication and characterisation; additional translocation datasets for DNA and NP/DNA complexes as well as example current-time traces and events for relevant event classes; characterisation results for the amplifier used as well as circuit simulations for translocation and insertion events; additional translocation frequency data in the presence and absence of a NP/DNA/NP construct.

## AUTHOR INFORMATION

**Cengiz Khan** – *University of Birmingham, School of Chemistry, Edgbaston Campus, Birmingham B15 2TT, United Kingdom*

**Oliver Irving** – *University of Birmingham, School of Chemistry, Edgbaston Campus, Birmingham B15 2TT, United Kingdom*

**Rand A. Al-Waqfi** - *Department of Medicinal Chemistry and Pharmacognosy, Faculty of Pharmacy, Jordan University of Science and Technology, P.O. Box 3030, Irbid 22110, Jordan*

**Giorgio Ferrari** - *Department of Physics, Politecnico di Milano, P.za L. da Vinci 32, Milano, 20133, Italy*

**Tim Albrecht** - *University of Birmingham, School of Chemistry, Edgbaston Campus, Birmingham B15 2TT, United Kingdom*

**Corresponding Author**

***Tim Albrecht** − University of Birmingham, School of Chemistry, Birmingham B15 2TT, U.K.; https://orcid.org/0000-0001-6085-3206; Email: t.albrecht@bham.ac.uk

**Present Addresses**

**Author Contributions**

The manuscript was written through contributions of all authors. All authors have given approval to the final version of the manuscript.

**Funding Sources**

Parts of this work were funded by the Leverhulme Trust (RPG-2022-165). RAAW acknowledges financial support from the Jordan University of Science and Technology.

**Notes**

**ACKNOWLEDGMENT**
**ABBREVIATIONS**

AC, Alternating Current; DBCO, Dibenzocyclooctyne; DC, Direct Current; NEMS, Nanoelectromechanical systems; NP, Nanoparticle; TEM, Transmission Electron Microscopy.

**TOC figure**

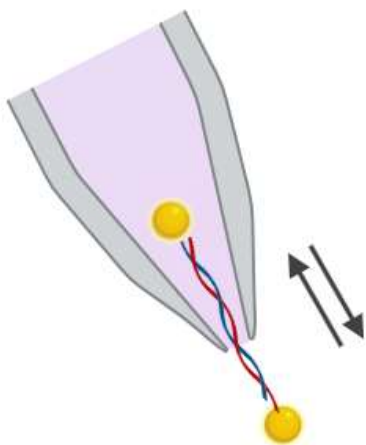

# Constructing a Nanopipette-based DNA Electro-Mechanical Device

*Cengiz J. Khan, Oliver J. Irving, Rand A. Al-Waqfi, Giorgio Ferrari, Tim Albrecht\**

University of Birmingham, School of Chemistry, Edgbaston Campus, Birmingham B15 2TT, United Kingdom

Supporting Information

**S1. Synthesis of end-functionalised 5 kbp DNA**

Primer sequences are provided in Appendix 1.1. A PCR reaction was assembled using: 2.5 μL of 10 μM forward primer, 2.5 μL of reverse primer (IDT), 1 μL of 1 ng/μL lambda DNA (Sigma Aldrich), 19 μL of nuclease-free water (VWR), and 25 μL of 2X Q5® Hot Start Hi-Fidelity Master Mix (New-England Biolabs). The mixture was prepared on ice and run on a PrimeG thermocycler (Cole Parmer) under cycling conditions listed in Appendix 1.2. The PCR product was purified using a Monarch DNA Cleanup Kit (New-England Biolabs).

DNA samples were analysed by agarose gel electrophoresis. A 1% gel was prepared by dissolving 1g agarose in 100 mL 1X TAE buffer (VWR), heating to 70°C, and pouring into a casting tray. Samples (5 μL DNA + 1 μL TriTrack loading dye (Thermofischer)) and 5 μL GeneRuler 1 kb ladder (Thermofischer) were loaded and run at 80 V for 45 minutes. Gels were stained with 1X SYBR Gold (Invitrogen) for 45 minutes and imaged using a UV Vis illuminator (GelDoc Go system, Bio-Rad). DNA concentrations was determined using a NanoDrop spectrophotometer (Schimadzu) by measuring absorbance at 260nm.

The sequences and PCR cycling parameters are shown in Table S1a and Table S1b respectively.

**Table S1a** - Primer Sequences

| Name | Sequence (5' – 3') | 5' Modification |
|---|---|---|
| Forward Primer | ATTTACAGCGGCAGCCATAAGGT | Biotin |
| Reverse Primer | TCATCAGGGCGAGATGCTCAATG | Azide |

**Table S1b** – PCR cycling parameters

| STEP | TEMP/°C | TIME/s |
|---|---|---|
| Initial Denaturation | 98 | 30 |
| 30 Cycles | 98 | 10 |
|  | 70 | 20 |
|  | 72 | 150 |
| Final Extension | 72 | 120 |
| Hold | 5 | - |

## S2. DNA-AuNP conjugation

Purified 5 kbp DNA was diluted to a final concentration of 11.6 nM in nuclease-free water. A 60 μL aliquot of this DNA was mixed with 60 μL of 11.6 nM DBCO-functionalised AuNPs (Nanopartz Inc.). The reaction was incubated at room temperature for 18 hours on a shaker to allow copper-free click conjugation via DBCO-azide linkage. The resulting reaction mixture was used directly for translocation and trapping experiments.

For the agarose gel and TEM imaging experiments (main text, fig. 2), the full construct was prepared in solution. For this purpose, 60 μL of 11.6 nM streptavidin-functionalised AuNPs (Nanopartz Inc.) were added to 60 μL of the DBCO-AuNPs/5 kbp DNA mixture. The reaction was incubated at room temperature for 30 minutes on a shaker to allow for biotin-avidin binding. This enabled the formation of both half constructs (DNA with a single AuNP) and full dumbbell structures (DNA tethered to two AuNPs), allowing further structural characterisation. More examples of TEM imaging of full dumbbell structures are shown in fig S1.

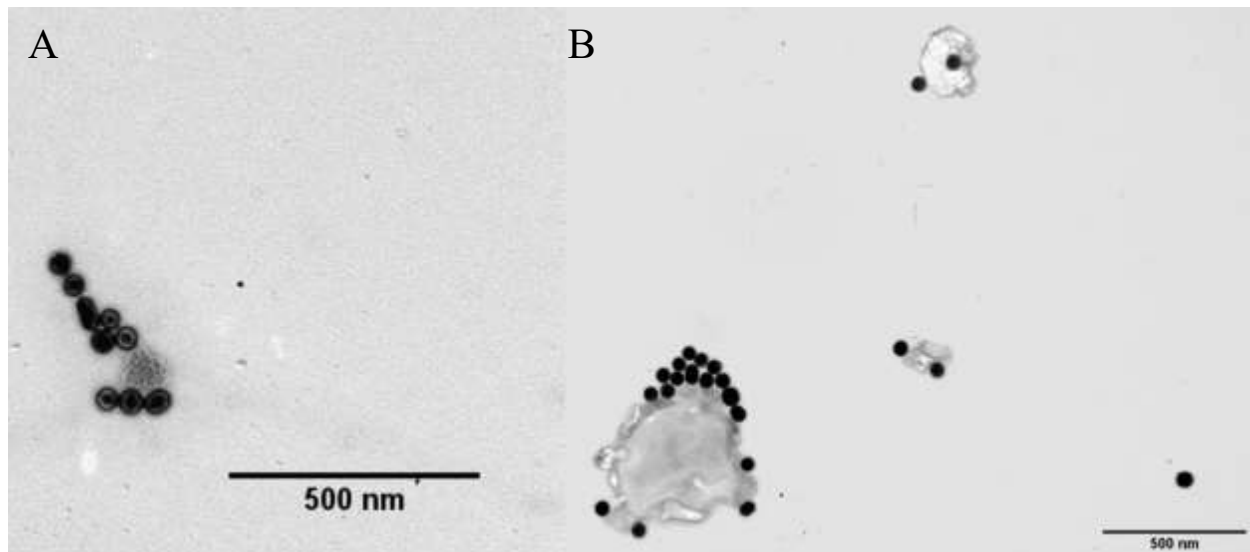

**Figure. S1** Further TEM imaging of NP/DNA/NP dumbbell structure mixtures acquired using JEOL JEM-1400 transmission electron microscope at 80 kV (Panels A and B). Dumbbell-like assemblies are visible among regions of particle clustering and DNA agglomeration, likely arising

from the drop-casting process. While the overall morphology remains well preserved, localized beam-induced degradation of the lighter DNA regions is observed in panel B.

### S3. Particle size distribution

A 1 µL aliquot of 0.1 nM AuNPs was drop-cast onto a copper TEM grid (Electron Microscopy Sciences, Pennsylvania, USA) and left to dry for 1 h under ambient conditions. The grid was subsequently washed with 10 µL of distilled water to remove residual impurities and dried for an additional 1 h. Imaging was performed using a JEOL 1400 transmission electron microscope, and representative micrographs are shown in fig. S1a. The micrograph was then analyzed using ImageJ to obtain particle size distribution as shown in fig. S2b.

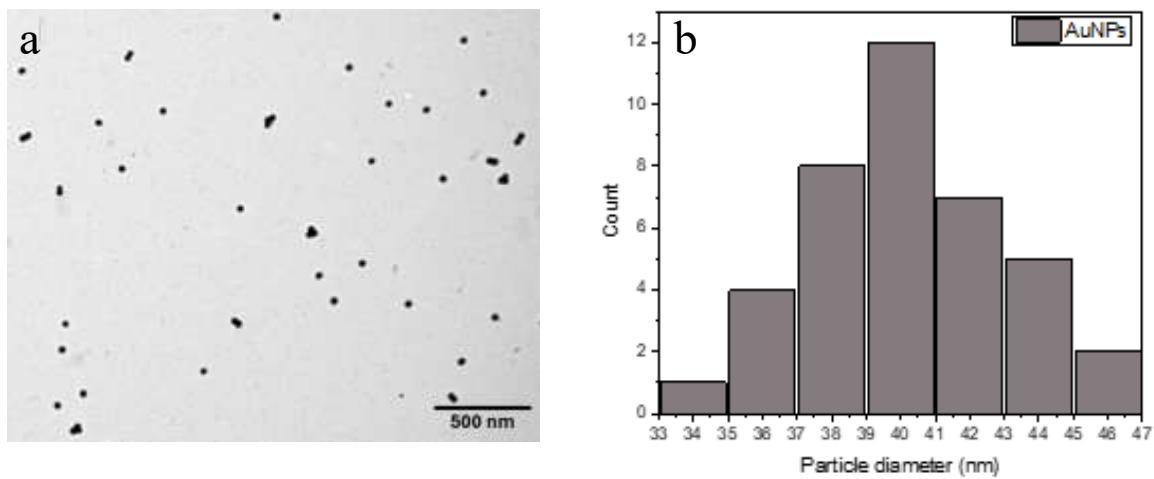

**Figure. S2** A sample of deposited AuNPs was imaged using a JEOL JEM-1400 transmission electron microscope at 60kV. Approximately 40 particles were manually measured (horizontally) using ImageJ software. The resulting histogram showed an approximately normal distribution centered at 41 nm ± 3.7 nm, consistent with vendor specifications of 40-42 nm.

## S4. Nanopipette fabrication and electrical characterisation

Quartz capillaries (1.00 mm OD, 0.50 mm $D_i$, 7.5 cm length; World Precision Instruments) were plasma-cleaned for 5 minutes and pulled into nanopipettes using a P-2000 laser puller (Sutter Instruments) with Program 59 (parameters in Table S2a). The resulting pipettes were imaged under a Nikon Eclipse Ti2 optical microscope to measure taper lengths of the nanopipettes. The taper length is defined as the length from the tip of the capillary to where the internal channel had reached $D_i$. Ag/AgCl electrodes were made by cutting 10 cm of silver wire (0.25 mm diameter, 99.99% purity, Goodfellow) and immersed in 38% v/v nitric acid (Sigma Aldrich) for 10 s, then washed with Milli-Q water (18 MΩ, Merck Millipore) to remove surface impurities. The cleaned wires were soldered to gold contact pins and submerged in 4 M LiCl 1xTE solution. Annodization was performed in an electrochemical cell using a gold wire (99.99% purity, Goodfellow) as a counter electrode and applying a current of 1 mA for 5 minutes, until the electrode surface turned black.

I-V measurements were performed using a CompactStat potentiostat (Ivium Technologies) to calculate the pore conductance, G. Assuming a known ionic conductivity, g(c) of 173 mS cm$^{-1}$ for 4 M LiCl, the pore diameter ($d_{pore}$) was estimated from G and taper length (l) using Equation 1.[1]

$$d_{pore} = \frac{4Gl + \frac{\pi}{2}GD_i}{D_i \pi g(c) - \frac{\pi}{2}G} \qquad (1)$$

A summary of pipette geometries and conductance-derived diameters is shown in Table S2b. From this set, pipette 1 was selected for the translocation of unbound dsDNA (control), while pipettes 2 and 6 were used for all device experiments.

**Table. S2a** P-2000 pull parameters (Programme 59)

| Programme 59 | | | | | |
|---|---|---|---|---|---|
| Line | Pull | Heat | Filament | Velocity | Delay |
| 1 | 75 | 700 | 5 | 35 | 150 |
| 2 | 200 | 700 | 0 | 15 | 128 |

**Table. S2b**

| Pipette | Taper length (µm) | Conductance (nS) | Pore diameter (nm) |
|---|---|---|---|
| 1* | 3060 | 38 | 19 |
| 2** | 3125 | 47 | 24 |
| 3 | 3400 | 26 | 14 |
| 4 | 3263 | 48 | 25 |
| 5 | 2834 | 74 | 33 |
| 6 | 3119 | 50 | 24 |
| 7 | 3324 | 27 | 14 |
| 8 | 2987 | 51 | 24 |
| 9 | 3245 | 40 | 18 |
| 10 | 3256 | 34 | 14 |
| Mean | 3161 | 44 | 21 |
| SD | 170 | 14 | 6 |

*Pipette used for unbound dsDNA control experiments

**Pipette used for DNA-AuNP device measurements.

## S5 DNA translocation and trapping experiments

DNA was injected separately into the bulk solution in a 3 mL liquid cell containing 2 mL 4 M LiCl 1XTE to a final concentration of DNA ~300 pM. The liquid cell was housed in a double Faraday cage to reduce electrical interference. A negative bias value means that the electrode outside the nanopipette is biased negatively, thereby resulting in an electrophoretic driving force for (negatively charged) DNA to translocate into the pipette.

Experiments were conducted in a semiautomated fashion using in-house MATLAB code with a sequence of applied biases, where for each bias, 102 data files of 10 s each were collected before the next bias value was applied.

Data recording was performed at a sampling rate of 1 MHz using a custom-built low-noise, high-bandwidth amplifier connected to the digital oscilloscope for analogue-to digital conversion (Picoscope 4262 Pico Technology), as reported previously.[1-4] Briefly, in this design, the input current is split into two output channels, namely the ''DC'' and ''AC'' channels. The former contains slow modulations of the input current (cutoff frequency ~ 7 Hz), including the open pore current. The AC channel contains fast modulations of the input current, for example, (short-lived) standard translocation events, and usually is zero mean, facilitating baseline correction.

Using custom-built MATLAB code, events were detected with a $5\sigma_{AC}$ threshold, where $\sigma_{AC}$ was the standard deviation of the noise in the AC channel. For each detected event, relevant segments of the current-time trace were extracted from and up to the adjacent zero crossings and relevant event characteristics were determined, such as the event duration based on $1\sigma$ threshold crossings ($\tau$), which we found to better capture the characteristics of events with complex shapes. Additionally, the effective current values ($\Delta I_e$) of the events were calculated by dividing the event charge deficit (q) by event duration ($\tau$). Where, event charge (q) is the integral of the current signal over the duration of the detected event[1, 3-5].

Finally, to calculate the change in average $I_{DC}$ ($\Delta I_{DC}$) and in the AC channel noise ($\Delta \sigma_{AC}$) we recorded the $I_{DC}$ and $\Delta \sigma_{AC}$ values at 100 ms time windows before and after the event ends in the AC channel.

## S6 Translocation NP-DNA and control DNA at different $V_{bias}$

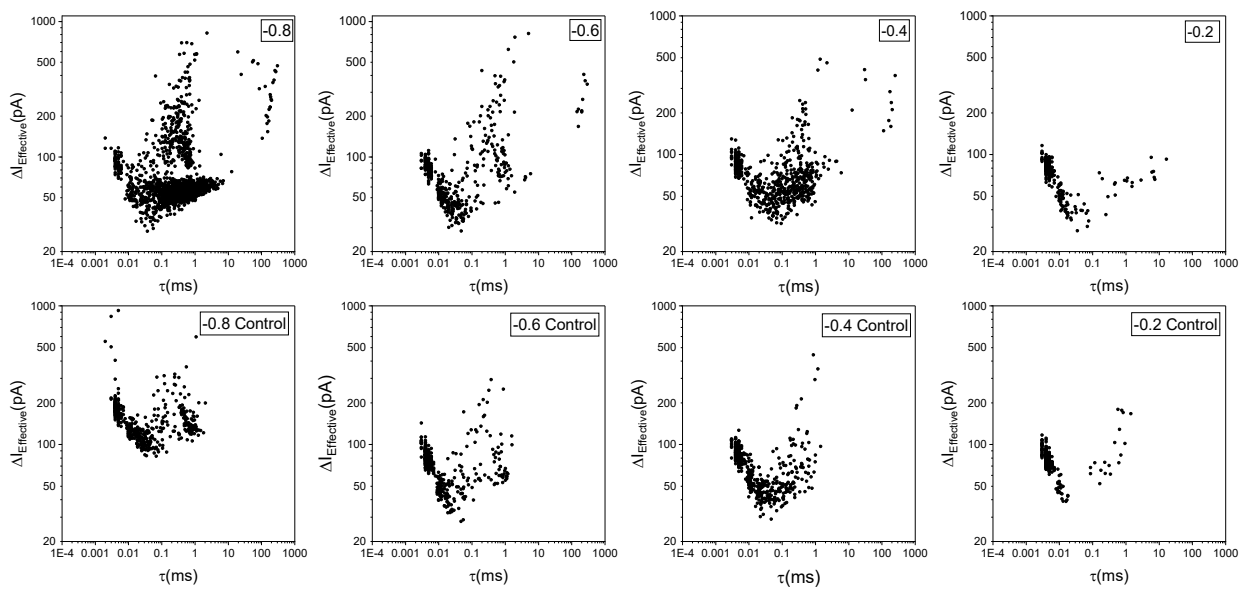

**Figure S3.** Scatter plots of $\Delta I_e$ vs. $\tau$ for AuNP–DNA mixtures and DNA-only control samples recorded across applied biases (±0.2 to ±0.8 V). Both datasets were obtained using 300 pM PCR-amplified DNA, with control experiments performed in the absence of AuNPs in the electrochemical cell. Each scatter plot corresponds to a continuous 1000 s recording prior to switching to the next bias condition.

# S7 Example current-time traces from translocation experiments

**Figure S4a.** Unbound dsDNA (control)

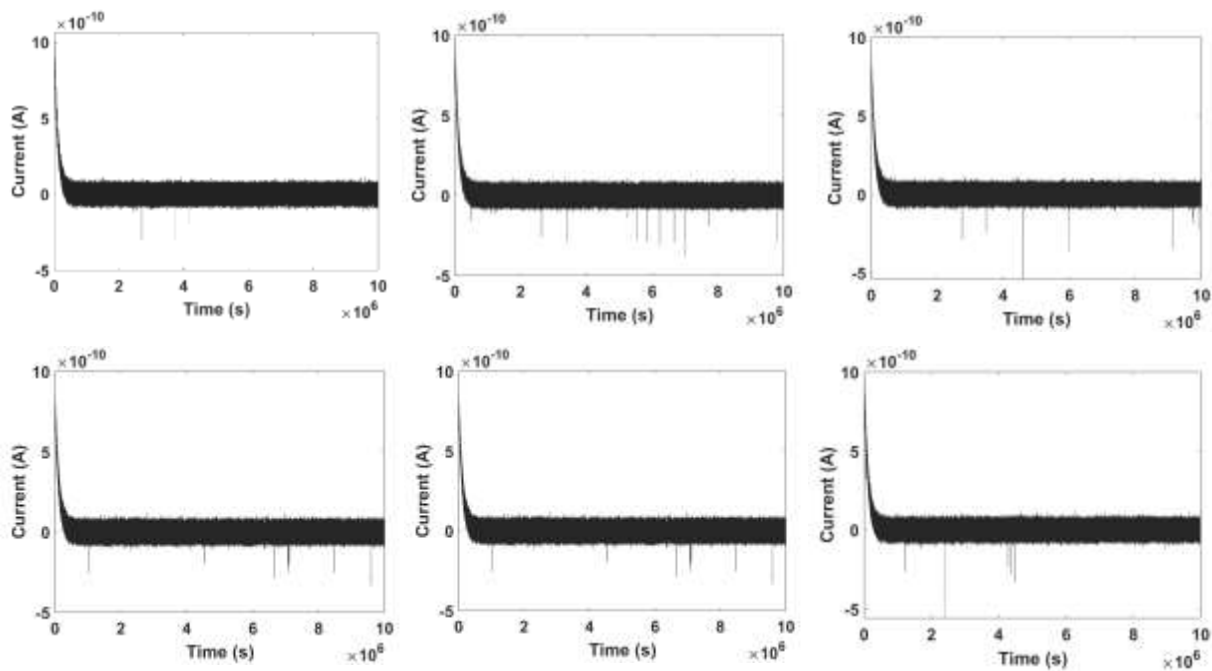

**Figure S4b** NP/DNA insertion experiment

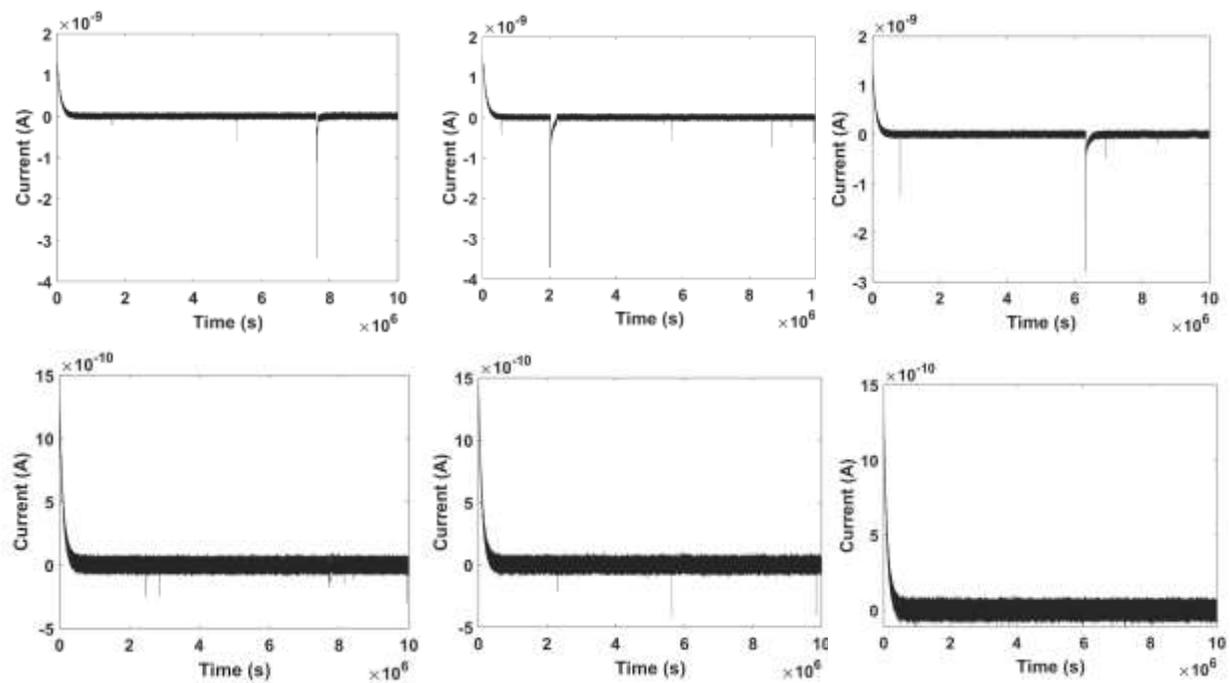

## S8 Example events

**Figure S5a** Linear DNA events

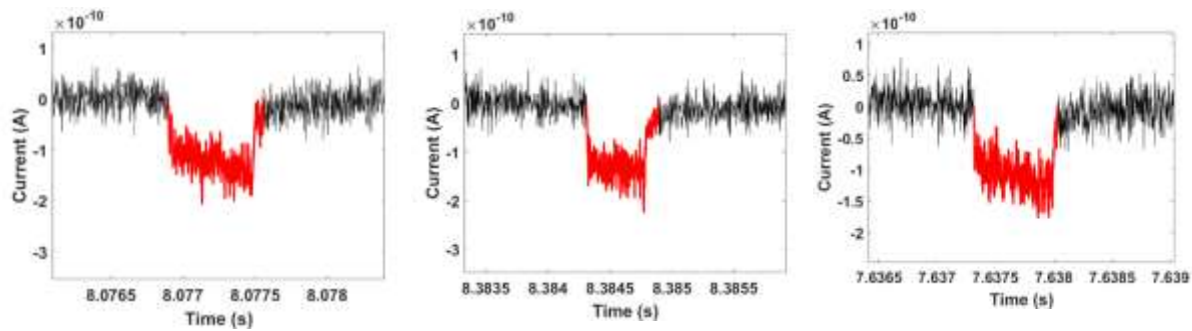

**Figure S5b** Non-linear (folded) DNA events

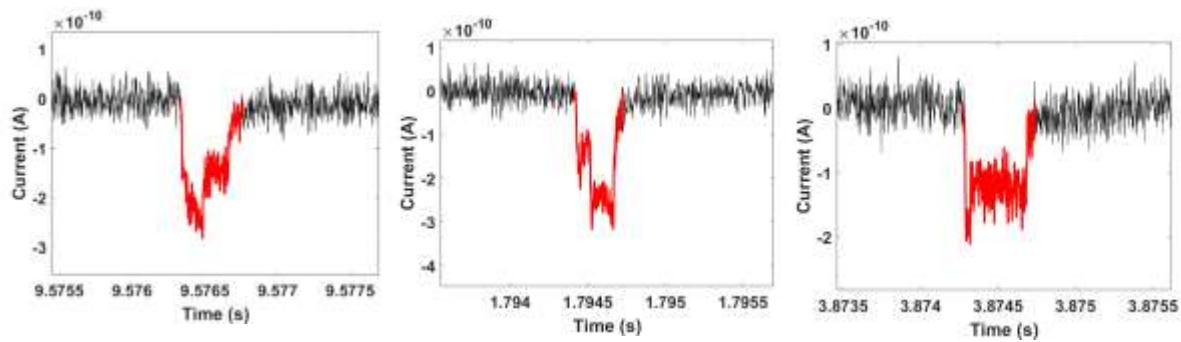

**Figure S5c** Intermittent insertion NP/DNA events

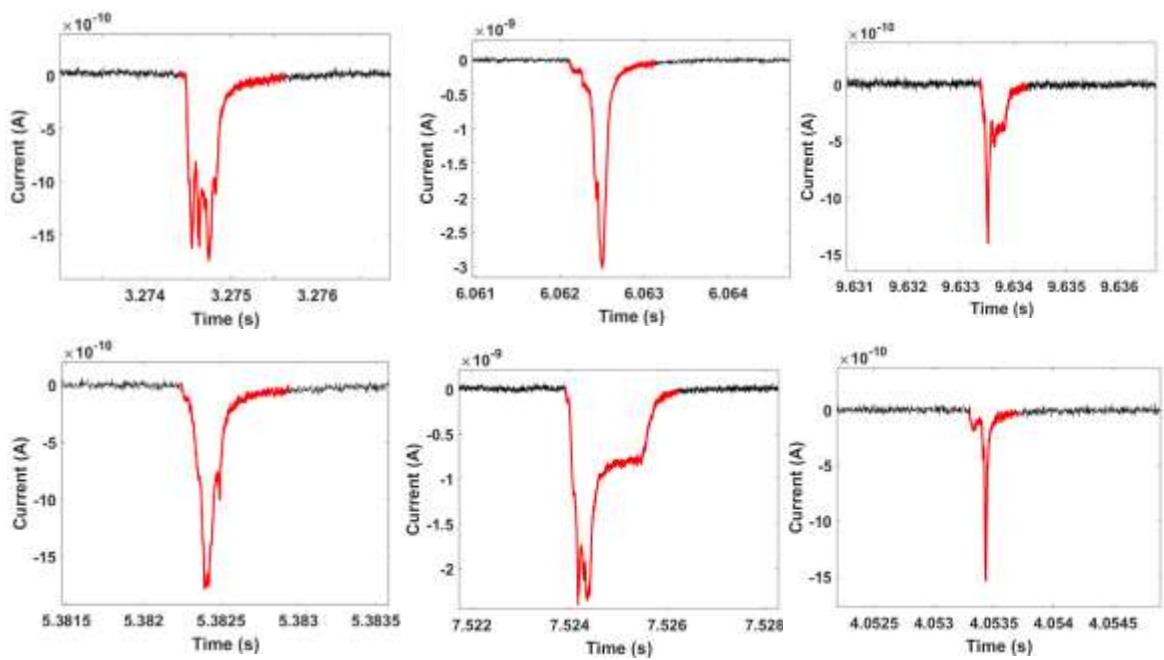

**Figure S5d** Stable insertion Events

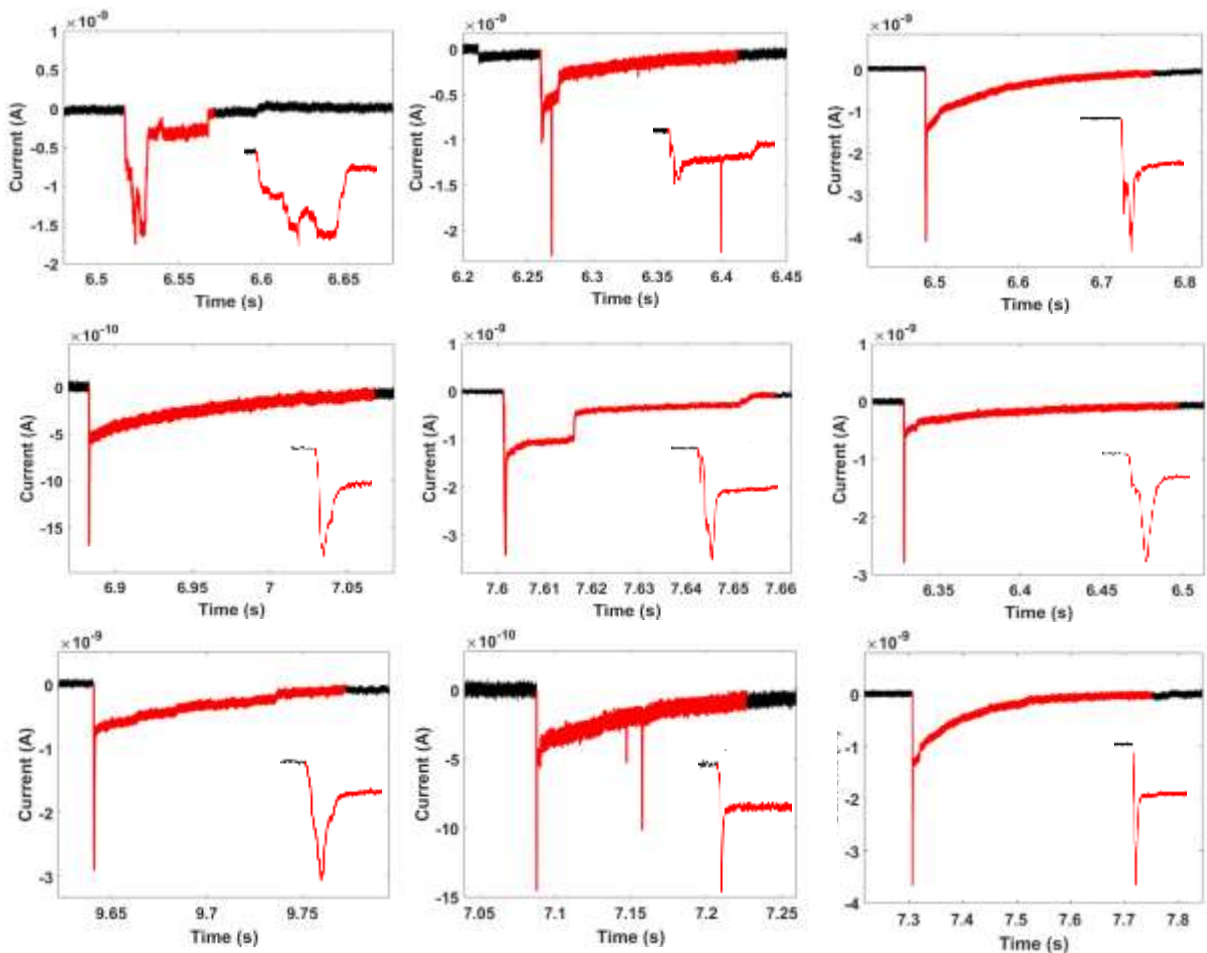

**Figure S5d.** Representative insertion events for AuNP–DNA at –0.8 V. Traces show current blockades exceeding 1 nA followed by a gradual AC modulation returning to baseline. Several events display sub-features consistent with initial particle–pore collisions, highlighted in the insets.

## S9 – CMOS current amplifier

The current $I_{IN}$ coming from the nanopipette is measured using the custom amplifier shown in fig. S6.

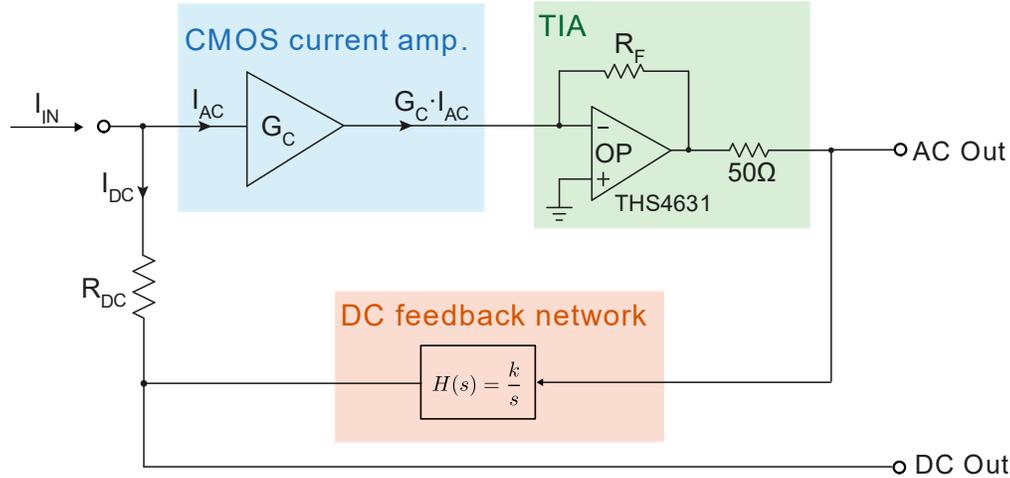

**Figure. S6**: Simplified scheme of the custom amplifier with two output channels.

It is based on a low-noise wide-bandwidth CMOS current amplifier with a gain $G_C$ = 990 and bandwidth of 1MHz.[6] The amplified current is converted into a voltage by a standard transimpedance amplifier with a feedback resistor of $R_F$ = 51 kΩ. Since the noise of the CMOS current amplifier is proportional to the DC current at its input, an additional feedback network H(s) operating at low frequencies forces the stationary input current into the off-chip resistor $R_{DC}$ = 100 MΩ. H(s) is made by an integrator stage (k ≅ 20 s$^{-1}$) whose gain decreases with frequency, deactivating the feedback loop at high frequency and allowing the current amplifier to amplify the fast input current variations. As a result, the voltage across the resistor $R_{DC}$ (DC output) is a low-pass filtered version of the input current $I_{IN}$ given by equation 2.

$$V_{out,DC} = \frac{R_{DC}}{1+s\tau_{DC}} I_{IN} \qquad (2)$$

where $\tau_{DC}= 1/2\pi f_{DC}= R_{DC} /( G_{DC} \cdot R_F \cdot k) \cong 22$ ms is the time constant of the feedback loop. On the contrary, the TIA output (AC output) is a high-pass filtered version of the input current spwm in equation 3:

$$V_{out,AC} = G_c\, R_F \frac{s\tau_{DC}}{(1+s\tau_{DC})(1+s\tau_H)}\, I_{IN} \qquad (3)$$

where $\tau_H$ is the time constant of the high-frequency pole ($\approx 2$ MHz) that limits the overall bandwidth of the custom amplifier and is given by the TIA.

The simulated step response of the amplifier is reported in Fig. S7 in the case of an abrupt variation from 1 nA to 800 pA. The outputs of the amplifier are reported divided by the nominal gains, i.e., $R_{DC}$ for the DC channel and $G_C \cdot R_F$ for the AC channel. As expected, the DC channel maintains the information on the mean value of the input current and follows the input variation with a settling time of about 100 ms. The AC channel correctly measures the fast variation of the input current on a time scale shorter than the response time of the DC feedback loop, $\tau_{DC}$. On a longer time scale, the input current is increasingly forced by the DC feedback network to flow in $R_{DC}$, reducing the AC output and correspondingly changing the DC output.

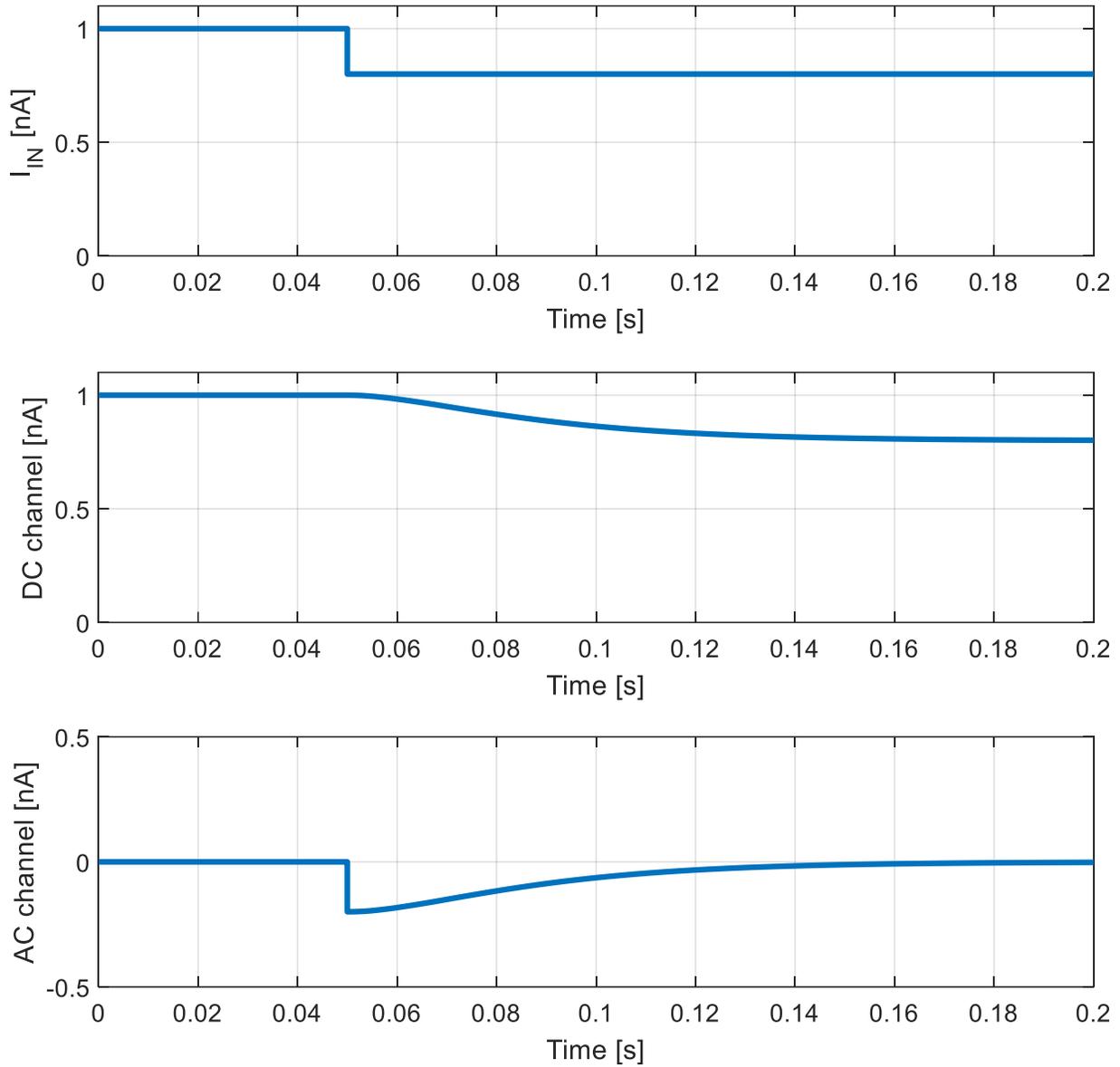

**Figure. S7**: Simulated response of the amplifier to a current step from 1nA to 800pA.

Fig. S8 reports the simulated response of the amplifier to a current pulse with a duration of 1 ms. Since the pulse duration is lower than the response time of the feedback loop, $\tau_{DC} \cong 22$ms, the voltage across $R_{DC}$ (i.e. the DC output) has no time to change significantly. Consequently, the current in $R_{DC}$ remains constant during the pulse, allowing the CMOS current amplifier and the TIA to fully process the current variation, as shown by the AC channel output.

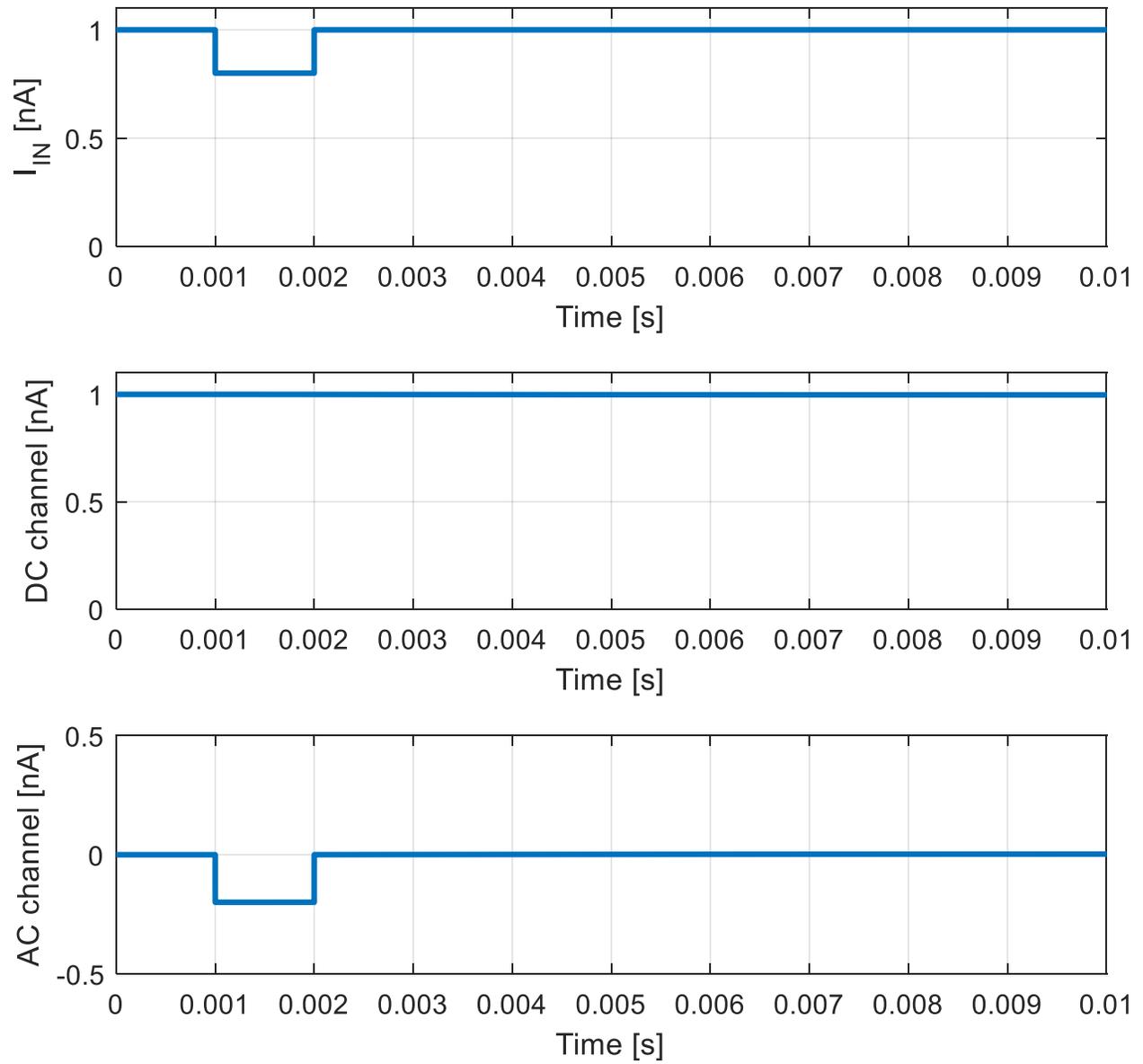

**Figure. S8**: Simulated response of the amplifier to a current pulse of 200pA and duration 1ms.

**S10. Comparison of translocation frequencies, in the presence and absence of NP/DNA/NP**

The translocation of free dsDNA continued in the presence of the NP/DNA/NP construct, albeit at a significantly reduced frequency. Fig. S9 below shows the probability distribution of translocation frequencies in the two cases. Even though the nanopore is smaller in the control experiment ($d_i \approx$ 19 nm vs. 24 nm), the mean translocation frequency is larger than for the nanopipette with the trapped NP/DNA/NP construct.

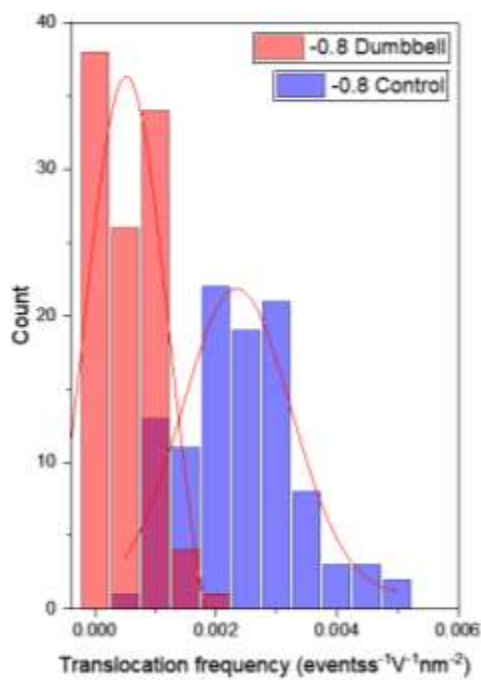

**Figure. S9** shows normalised translocation frequencies of 5 kbp DNA for two conditions: with the trapped NP/DNA/NP structure (red bars, $d_i \approx$ 24 nm), and control (blue bars, $d_i \approx$ 19 nm).